\documentclass[twocolappendix,twocolumn]{aastex631}
\usepackage[T1]{fontenc}
\usepackage{ae,aecompl}
\usepackage{bethmacros}
\usepackage{graphicx}
\usepackage{amssymb}
\usepackage{amsmath}
\usepackage{placeins}
\usepackage{times}

\def\spose#1{\hbox to 0pt{#1\hss}}
\def\lta{\mathrel{\spose{\lower 3pt\hbox{$\mathchar"218$}}
     \raise 2.0pt\hbox{$\mathchar"13C$}}}
\def\gta{\mathrel{\spose{\lower 3pt\hbox{$\mathchar"218$}}
     \raise 2.0pt\hbox{$\mathchar"13E$}}}

\newcommand{\appropto}{\mathrel{\vcenter{
  \offinterlineskip\halign{\hfil$##$\cr
    \propto\cr\noalign{\kern2pt}\sim\cr\noalign{\kern-2pt}}}}}

\shorttitle{Probing Feedback with Stellar Halo Outskirts}
\shortauthors{B.W. Keller}

\begin{document}
\title{Where Did the Outskirts Go? Outer Stellar halos as a
Sensitive Probe of Supernova Feedback}
\author[0000-0002-9642-7193]{B.W. Keller}
\affiliation{Department of Physics and Materials Science, University of Memphis,
\\ 3720 Alumni Avenue, Memphis, TN 38152, USA}
\email{bkeller1@memphis.edu}
\begin{abstract} 
    A recent comparison by \citet{Merritt2020} of simulated and observed Milky
    Way-mass galaxies has identified a significant tension between the outskirts
    ($r>20\kpc$) of the stellar halos in simulated and observed galaxies.  Using
    observations from the Dragonfly telescope and simulated galaxies from the
    Illustris-TNG100 project, \citet{Merritt2020} finds that the outskirts of
    stellar halos in simulated galaxies have surface densities $1-2$ dex higher
    than observed galaxies. In this paper, we compare two suites of 15 simulated
    Milky Way-like galaxies, each drawn from the same initial conditions,
    simulated with the same hydrodynamical code, but with two different models
    for feedback from supernovae.  We find that the MUGS simulations, which use
    an older ``delayed-cooling'' model for feedback, also produce too much
    stellar mass in the outskirts of the halo, with median surface densities
    well above observational constraints.  The MUGS2 simulations, which instead
    use a new, physically-motivated ``superbubble'' model for stellar feedback,
    have $1-2$ dex lower outer stellar halo masses and surface densities. The
    MUGS2 simulations generally match both the median surface density profile as
    well as the scatter in stellar halo surface density profiles seen in
    observed stellar halos.  We conclude that there is no ``missing outskirts''
    problem in cosmological simulations, provided that supernova feedback is
    modelled in a way that allows it to efficiently regulate star formation in
    the low-mass progenitor environments of stellar halo outskirts.    
\end{abstract}


\section{Introduction}
The stellar halos of galaxies provide an interesting laboratory for testing
theories of cosmological galaxy formation.  The long dynamical times
\citep[e.g.][]{Bullock2005} and high stellar ex-situ fractions
\citep[e.g.][]{Font2011} outside of the stellar disk can allow the structures
formed by the accretion and tidal disruption \citep{Bullock2005} of progenitors
to persist for a significant fraction of a Hubble time, preserving a relic of
the accretion history of the galaxy.  However, the low stellar mass (relative to
the disk) of stellar halos \citep{Merritt2016} makes them difficult to resolve
in cosmological simulations, and their extremely low surface brightness makes
them exceptionally difficult to observe.  Some of the most heavily studied
issues in galaxy formation over the past decades have related to the abundance
and luminosity of satellites within the halo of Milky Way (MW)-like galaxies.
The ``missing satellites'' problem \citep{Klypin1999} points to the
significantly larger number of satellite galaxies seen in dark-matter only
simulations of MW-like galaxies compared to the observed satellite mass
functions of the MW and M31.  The ``Too-Big-To-Fail'' problem
\citep{Boylan-Kolchin2011} heightens this tension by showing that CDM predicts
the MW halo should contain satellites much more massive or centrally
concentrated (with $V_{max}>30\kms$) than what has been measured for the
kinematics of the brightest MW satellites.  Many solutions to these problems
have been proposed, from reionization regulating star formation in low-mass
dwarfs \citep{Bullock2000,Gnedin2000} to enhanced tidal disruption by the
stellar disk \citep{Kelley2019}.  The advent of higher-resolution cosmological
simulations, using better models for ``strong'' stellar feedback
\citep[e.g.][]{Okamoto2010,Nickerson2013,Sawala2016,Akins2021} has produced a
growing consensus that the issues of satellite abundances around MW-like
galaxies is largely solved when the full range of baryonic processes (star
formation and feedback primarily) are accounted for.  Recently, a comparison
between a new survey of stellar halos around MW-mass galaxies and the
Illustris-TNG100 cosmological simulations \citep{Merritt2020} has pointed
towards a tension between observed stellar halos and simulated stellar halos.
\citet{Merritt2020} has found the stellar halos in Illustris-TNG100 simulated
galaxies have significantly higher surface densities at radii $r>20\kpc$
compared to observations from \citet{Merritt2016}, a problem the authors dub
``missing outskirts''.

The most precise observations of stellar halos are those of our own galaxy and
of Andromeda (M31).  For these two L* galaxies, we have detailed censuses of the
positions \citep[e.g.][]{Searle1978}, ages \citep[e.g.][]{Helmi1999},
metallicities \citep[e.g.][]{Ibata2001}, and kinematics
\citep[e.g.][]{Yanny2003} for thousands of individual stars within the stellar
halo (now tens of thousands with the advent of Gaia).  However, surveys of other
galaxies have suggested that both the MW and M31 may have relatively extremal
stellar halos, with M31 having an unusually massive, iron-rich stellar halo
compared to the MW's unusually light , iron-poor stellar halo (see Figure 12 of
\citep{Harmsen2017}, but note that this result is not seen in the DNGS galaxies
examined by \citealt{Merritt2016}).  The MW may also have a more elliptical
stellar halo than galaxies with similar stellar mass \citep{DSouza2014}.
\citet{Deason2013} has also concluded that the shapes of the MW and M31 stellar
halo profiles suggest a prolonged accretion history for M31 and an accretion
history dominated by one or more massive satellites in the MW. Surveys of
extragalactic stellar halos have been primarily limited by the low surface
brightness of stellar halos, typically 8-10 $\rm{mag\; arcsec}^{-2}$ lower than
the stellar disk.  Studies of stellar halos around other galaxies have typically
relied on either stacking \citep[e.g.][]{DSouza2014,Wang2019} or on deep,
``pencil-beam'' studies that observe only a small fraction of the covering area
of a galaxy's halo (The GHOSTS Survey \citep{RadburnSmith2011}, for example,
used the Hubble Space Telescope's Advanced Camera for Surveys to examine the
halos along the major and minor axes of 14 nearby disk galaxies).  The advent of
telephoto array telescopes, such as Dragonfly \citep{Abraham2014} and Huntsman
\citep{Spitler2020} provide two critical features that make them ideal
instruments for observing stellar halos.  The first, a feature of the advanced
anti-reflection coatings available in modern high-end photography lenses, is
extremely low imaging noise, giving them the ability to probe surface
brightnesses below $\sim30\;\rm{mag\; arcsec}^{-2}$.  The second, owing to their
construction from consumer/off-the-shelf photography equipment, is a large field
of view ($\sim10\;\rm{deg}^2$), which allows them to cover the entire halo out
to nearly a virial radius for most nearby L* galaxies.  These new telescope
systems have led to a revolution in low surface brightness observations, and
have discovered new classes of unusual, low surface brightness galaxies
\citep{vanDokkum2015,vanDokkum2018}.

The Dragonfly Nearby Galaxy Survey (DNGS) \citep{Merritt2016} imaged 8 nearby,
MW-like galaxies (M101, NGC 1042, NGC 1084, NGC 2903, NGC 3351, NGC 3368, NGC
4220, and NGC 4258).  These observations probe down to a surface brightness of
$31\;\rm{mag\; arcsec}^{-2}$, measuring the stellar halo profiles and halo mass
fractions.  In their results, the authors found the stellar halo fractions beyond
5 half-mass radii $R_h$ were detectably lower than predictions for the stellar
halo mass from the Aquarius simulations \citep{Cooper2010}, the Eris simulation
\citep{Pillepich2015}, and the Millenium II simulation \citep{Cooper2013}.
Measuring stellar halo fractions is a somewhat tricky matter, as
\citet{Sanderson2018} demonstrated using simulations from the FIRE-2 and ELVIS
suites.  Because of the steep power-law density profile of stellar halos (with
density slopes of $-2$ to $-4$, \citealt{Font2011}), the choice of inner cut for
integrated mass measurements of the stellar halo can have a significant impact
on the total stellar halo mass and halo mass fractions: if the surface
density has a slope steeper than $-2$, most of the mass will be contained in the
inner halo.  \citet{Merritt2020} sidesteps this issue by comparing the radial
profiles of stellar surface brightness and mass surface density of the DNGS
galaxies to those of simulated
galaxies from the Illustris-TNG project \citep{Pillepich2018}.  These results
show that the outer stellar halos of simulated galaxies, beyond radii of
$\sim20\kpc$, are systematically higher in the entire population of simulated
galaxies, despite a careful ``apples-to-apples'' comparison of mock observations
to the actual DNGS data.

While sparse, observational data for stellar halos in MW-mass galaxies has still
been a useful tool for comparison to semi-analytic models and numerical
simulations of galaxy formation.  These comparisons have sought to test our
predictions on the assembly of stellar halos in the $\Lambda$CDM framework.
Many early models use analytic or semi-analytic models, combined with ``particle
tagging'' approaches applied to collisionless N-body simulations
\citep[e.g.][]{Bullock2005,Cooper2010,Cooper2017}.  In these approaches, dark
matter (DM) particles are tagged as carrying stellar mass into the galaxy halo,
based on their kinematic and host halo properties \citep[e.g.][]{Cooper2010}.
Recently, the improving resolution and physical fidelity of large, cosmological
galaxy simulations have allowed the study of stellar halos from self-consistent
models of galaxy formation, including both hydrodynamic effects and radiative
processes, as well as star formation and stellar feedback
\citep[e.g.][]{Font2011,Pillepich2014,Sanderson2018,Monachesi2019,Obreja2019b,Font2020}.
Different simulation studies of stellar halos have all relied on different
models for star formation and stellar feedback, a wide range of numerical
resolutions, and different simulation volumes.  Thus, comparing different models
to observations can allow us to tease out the impact these various choices and
assumptions have on the assembly of different stellar populations within the
galaxy.  

Simulation studies of the assembly of the outer stellar halos in MW-mass
galaxies have generally converged on a number of properties of the stellar
outskirts. Studies examining the stellar halos of the GIMIC \citep{Font2011},
Aquarius \citep{Tissera2012}, Eris \citep{Pillepich2015}, Auriga
\citep{Monachesi2019}, and ARTEMIS \citep{Font2020} simulated galaxies all find
that the outer halo is dominated by accreted (ex-situ) stars, while the inner
halo is composed of a mix of in-situ and ex-situ stars.  This difference
manifests itself in distinct changes in the chemical
\citep[e.g.][]{Font2011,Tissera2012,Pillepich2014} and kinematic
\citep[e.g.][]{Tissera2013,Cooper2015} properties as one moves out through the
stellar halo.  The density profile of the halo may be steeper for the outer halo
versus the inner halo\citep{Font2011,Pillepich2014,Font2020}, though other
studies \citep{Cooper2015,Thomas2021} have found a roughly constant slope out to
$\sim R_{vir}$.  For the remainder of this paper, we use ``stellar halo
outskirts'' and ``outer stellar halo'' to refer to the portion of the stellar
halo beyond $20\kpc$, following the division used by \citet{Carollo2010},
\citet{Pillepich2015}, and \citep{Merritt2020}.

In this paper, we examine the specific question raised by \citet{Merritt2020}:
do simulated galaxies systematically overproduce stars in the outskirts of the
stellar halo?  We specifically examine one critical piece of all modern galaxy
formation simulations, namely stellar feedback from core-collapse supernovae
(SNe).  Using a set of paired simulations of 15 MW-like galaxies, we compare how
two different numerical implementations of SNe feedback change the outskirts of
the stellar halos.  We demonstrate that the surface density and total mass of
stellar halo outskirts are highly sensitive to the model of SN feedback used to
simulate the galaxy.  In particular, we demonstrate that the impact of feedback
on stellar halos is much stronger than the impact on the total stellar mass,
which owes to the different assembly history of halo stars versus the stars that
make up the majority of the stellar mass in the disk of the galaxy.
\section{Methods}
\subsection{Simulations}

We compare matched pairs of cosmological zoom-in simulations of 15 MW-like L*
galaxies with observations of 8 galaxies from the DNGS.  Each of the 15 pairs of
galaxies is simulated from the same initial conditions (ICs), derived from the
McMaster Unbiased Galaxy Simulations (MUGS) project \citep{Stinson2006}.  These
simulations use a WMAP3 $\Lambda$CDM cosmology with $H_0 = 73\kms\Mpc^{-1}$,
$\Omega_m=0.24$, $\Omega_\Lambda=0.76$, $\Omega_b=0.04$, and $\sigma_8=0.76$
\citep{Spergel2007}.  The DM resolution of the simulations is
$1.1\times10^6\Msun$ per particle, while the initial gas particle mass is
$2.2\times10^5\Msun$.  All particles use a uniform softening length of
$312.5\pc$.

The pairs of simulations we compare were run as part of the MUGS and MUGS2
simulation projects (\citealt{Stinson2010} and \citealt{Keller2016}
respectively).  They share identical ICs and parameters used in
the gravity solution, but feature a number of differences in the hydrodynamic
solver and the sub-grid prescription for star formation and feedback.  The
original MUGS galaxies used a ``traditional'' pressure-density prescription for
Smoothed-Particle Hydrodynamics (SPH) in the code {\sc Gasoline}, which has
subsequently been found to artificially suppress mixing in shearing and
turbulent flows \citep{Agertz2007}.  MUGS2 is the first set of cosmological
simulations produced using the new {\sc Gasoline2} code \citep{Wadsley2017}.
{\sc Gasoline2} includes a treatment for turbulent diffusion
\citep{Wadsley2008}, improved time-step calculation \citep{Saitoh2009}, and a
new treatment for the calculation of $P\rm{d}V$ forces \citep{Wadsley2017}.
While the hydrodynamics improvements in MUGS2 result in a more accurate
treatment of the interstellar medium (ISM) and circumgalactic medium (CGM), the
primary differences arise due to the difference in sub-grid feedback models
between MUGS and MUGS2.

In the original MUGS sample, SN feedback is modelled using a
delayed-cooling ``blastwave'' mechanism \citep{Stinson2006}.  This model is
designed to prevent numerical overcooling by disabling radiative cooling in
feedback-heated gas until the end of the momentum-conserving snowplow phase.
Alternatively, in the MUGS2 simulations, SN feedback is modelled
using the \citet{Keller2014} ``superbubble'' model.  In this model, radiative
cooling is included self-consistently for SN-heated gas by temporarily including
a two-phase component within recently heated gas particles.  The two isobaric
phases each calculate their cooling rates using a separate density and
temperature, and mass is evaporated from the cold ``shell'' phase into the hot
``bubble'' phase using the classical evaporation rates derived in
\citet{Cowie1977}.  Once the entirety of a two-phase particle's cold shell is
evaporated, it can continue evaporating neighbours using a stochastic
evaporation method described in \citet{Keller2014}.  The difference in strategy
between these two models can be summarized as: ``blastwave'' feedback seeks to
prevent numerical overcooling by disabling cooling for some appropriate time,
while ``superbubble'' feedback prevents it by heating only the correct amount of
gas. In \citet{Keller2014} and \citet{Keller2015}, we demonstrated that
superbubble feedback drives significantly more mass-loaded galactic outflows
than the blastwave model, and can reduce the overall star formation rate and
bulge fraction in both isolated and cosmological simulations of MW-like galaxies.

We list in table~\ref{MUGS_table} the halo mass $M_{200}$, virial radius
$R_{200}$, stellar mass $M_*$ within the inner $30\kpc$ of the halo (rather than
total stellar mass, in order to match \citet{Merritt2020}), and stellar
half-mass radii $R_{*,1/2}$ for each of the galaxies in the MUGS and MUGS2
simulations.  As this table shows, below $M_{200}\sim~10^{12}\Msun$, the MUGS2
galaxies produce significantly less stellar mass within the inner $30\kpc$ than
the MUGS simulations, while above this mass the two simulation sets have similar
stellar masses (to within a factor of $\sim2$).  The MUGS2 galaxies above
$M_{200}~10^{12}\Msun$ also have a more massive stellar bulge, which in turn
lowers their half-mass radius by a factor of $0.5-4$.  In all instances here, we
define the virial radius $R_{200}$ (and the enclosed virial mass $M_{200}$) as
the radius around the halo such that the enclosed density is 200 times the
critical density ($\rho=200\rho_{\rm crit}$). In both the MUGS and MUGS2
simulations, halos were found in each of the simulations using the Amiga Halo
Finder \citep[AHF;][]{Knollmann2009}.  While slight ($\sim10\%$) differences in
$M_{200}$ and $R_{200}$ arise from the different feedback models used, these
differences do not significantly impact the results shown here.

\begin{table*}
    \begin{flushleft}
    \begin{tabular}{l|lllllllll}
        \hline
        Galaxy & $M_{200}^{MUGS}$ & $M_{200}^{MUGS2}$ & $R_{200}^{MUGS}$ & $R_{200}^{MUGS2}$
        & $M^{MUGS}_{*}$ & $M^{MUGS2}_{*}$ &
        $R^{MUGS}_{*,1/2}$ & $R^{MUGS2}_{*,1/2}$ \\
        \hline
        \hline
         g7124  & 38.1 & 36.6 & 145 & 143 & 3.46 & 0.5 & 1.37 & 4.72 \\
         g5664  & 44.9 & 47.7 & 153 & 157 & 4.15 & 0.9 & 1.57 & 3.81 \\
         g8893  & 53.5 & 58.0 & 163 & 167 & 5.73 & 0.7 & 1.82 & 8.18 \\
         g1536  & 64.1 & 64.9 & 173 & 174 & 7.95 & 1.8 & 1.74 & 6.53 \\
         g21647 & 70.2 & 74.4 & 178 & 181 & 5.81 & 1.5 & 4.25 & 3.02 \\
         g422   & 75.7 & 76.2 & 182 & 183 & 4.41 & 1.0 & 2.94 & 7.10 \\
         g22437 & 76.1 & 85.2 & 183 & 190 & 6.27 & 9.0 & 1.76 & 0.68 \\
         g22795 & 78.0 & 85.2 & 185 & 190 & 6.01 & 10.6 & 1.52 & 0.94 \\
         g3021  & 88.7 & 97.8 & 193 & 199 & 8.45 & 3.6 & 3.60 & 4.11 \\
         g28547 & 93.6 & 98.5 & 196 & 200 & 5.74 & 1.5 & 53.75 & 6.57 \\
         g24334 & 97.3 & 102.2 & 199 & 202 & 6.57 & 2.3 & 6.35 & 5.89 \\
         g4145  & 103.1 & 119.5 & 202 & 213 & 8.61 & 15.0 & 2.25 & 1.19 \\
         g25271 & 108.7 & 125.5 & 206 & 216 & 9.51 & 15.5 & 1.73 & 1.21 \\
         g15784 & 121.3 & 131.1 & 214 & 219 & 9.10 & 12.9 & 2.83 & 1.53 \\
         g15807 & 190.3 & 203.2 & 249 & 254 & 12.83 & 20.9 & 4.00 & 1.03 \\
    \end{tabular}
    \end{flushleft}
     \caption{Basic properties of the MUGS and MUGS2 galaxies compared in this
     study.  We show halo masses $M_{200}$, virial radii $R_{200}$,  stellar
     masses $M_*$ (within the inner $30\kpc$ of the halo), and stellar half-mass
     radii $R_{*,1/2}$ for each galaxy.  As can be seen, below a halo mass of
     $M_{200}\sim10^{12}\Msun$, the MUGS2 galaxies produce significantly less
     stellar mass than the MUGS galaxies.  These galaxies also tend to be less
     compact, with stellar half-mass radii typically more than twice as large as
     the MUGS equivalent.  The large outlier in stellar radii for the MUGS
     example of g28547 is due to a variation in the timing of a major merger
     that occurs slightly earlier in MUGS2 than in MUGS.  All mass units are
     given in $10^{10}\Msun$, and all length units are given in $\kpc$.}
    \label{MUGS_table}
\end{table*}
\begin{figure}
    \includegraphics[width=0.5\textwidth]{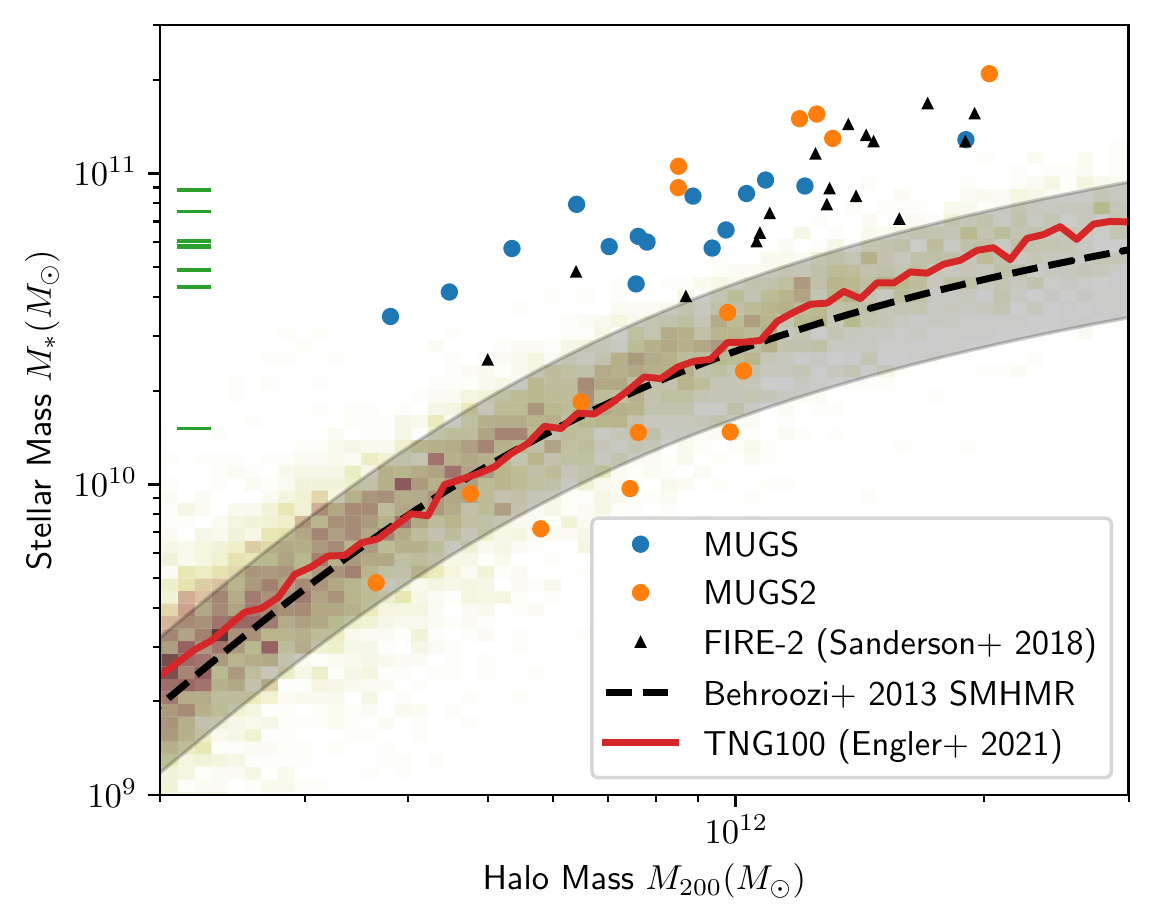}
    \caption{Stellar mass to halo mass relation in the MUGS, MUGS2, and DNGS
    galaxies.  DNGS does not measure halo masses, but the stellar masses are
    shown with the green dashes along the left side of the panel.  The
    abundance-matched stellar to halo mass relation from \citet{Behroozi2013} is
    shown as the dashed black curve, with one-sigma uncertainties in the grey
    region.  Black triangles show values for the FIRE-2 simulations examined by
    \citet{Sanderson2018}.  The 2D histogram shows the central
    galaxies in TNG100, with the red curve showing the average SMHMR for TNG100
    central galaxies \citep{Engler2021}.}
    \label{SMHMR}
\end{figure}
\begin{figure}
    \includegraphics[width=0.5\textwidth]{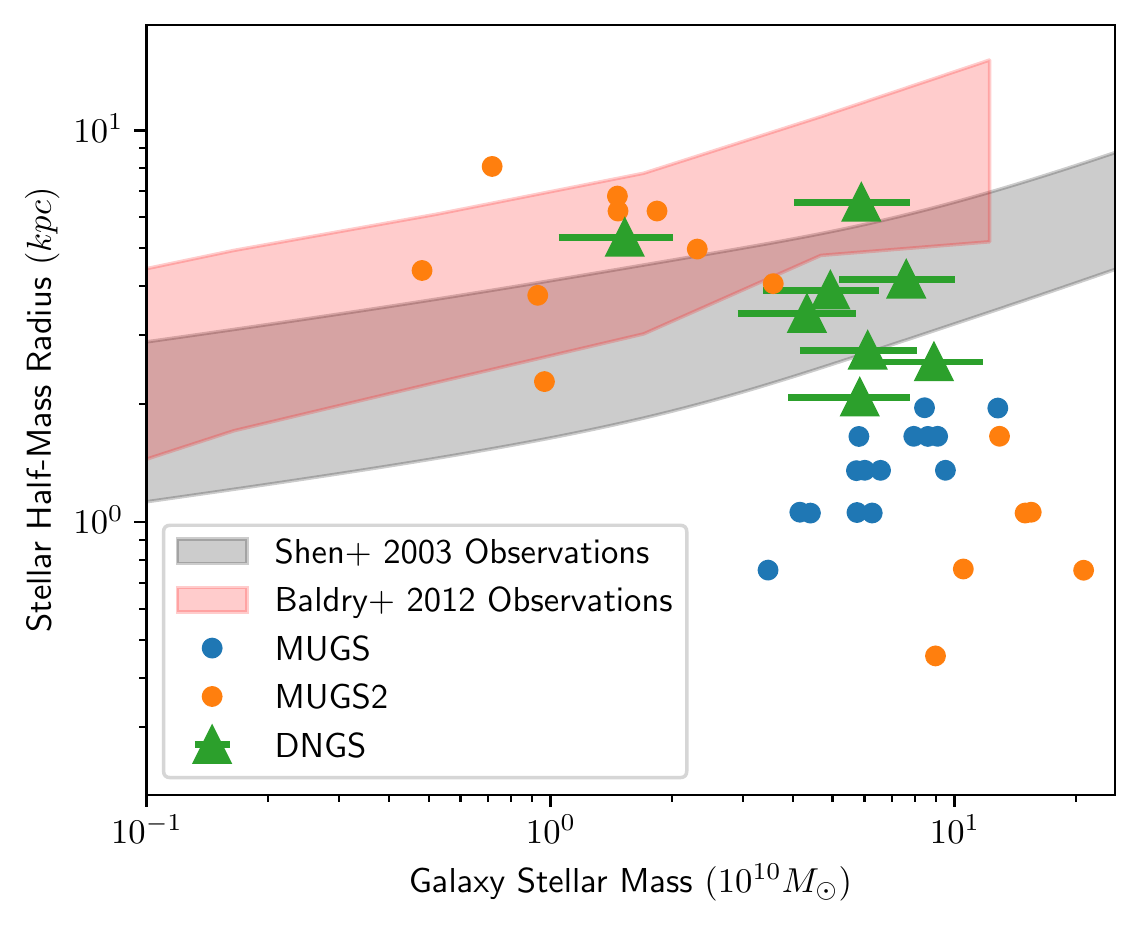}
    \caption{Stellar half-mass radius versus stellar masses for the MUGS, MUGS2,
    and DNGS galaxies.  The shaded regions show the fits to observed galaxy
    sizes from the SDSS \citep{Shen2003} and GAMA \citep{Baldry2012} surveys, in
    grey and red respectively.  As can be seen, for the more massive MUGS2
    galaxies (and all of the MUGS galaxies), the runaway growth of the stellar
    bulge results in half-mass radii much lower than the observed galaxy
    size-mass relation (and indeed, the typical sizes of the DNGS galaxies).}
    \label{size_mass}
\end{figure}

\subsection{Stellar Surface Density Maps}
\citet{Merritt2020} considers a number of potential observational and numerical
effects in comparing observations to mock stellar maps from the Illustris-TNG simulations
\citep{Pillepich2018}.  Simply radially binning star particles will of course not
allow a comparison to 2D image maps, and will suffer from discreteness noise in
the lower-density regions.  Discreteness noise is somewhat less important for
the higher baryonic mass resolution MUGS simulations we analyse here
($1.4\times10^6\Msun$ in TNG100 vs.  $2.2\times10^5\Msun$ in MUGS/MUGS2), but
for purposes of comparison we will follow the strategy of \citet{Merritt2020}.
Using the built-in analysis features of {\sc pynbody} \citep{pynbody}, we
generate smoothed 2D column density maps of stellar mass, using a cubic spline
kernel, with an adaptive smoothing length set by the distance to the 64th
neighbour particle.  These methods are designed to mimic the SPH smoothing
operations, and produce smoothed maps of quantities comparable to those analysed
in \citet{Merritt2020}.  We use smoothed maps with pixel dimensions of
$0.5\kpc$, as is used in \citet{Merritt2020}.  To further follow
\citet{Merritt2020}, we centre our galaxies at the deepest point in each halo's
potential well, but do not rotate the galaxies to orient the disk either edge or
face on, giving each galaxy an effectively random orientation.  For a brief
discussion of the (non-) importance of galaxy orientation to our results, see
appendix~\ref{orientation_robustness}.

\section{Results}
\begin{figure*}
    \includegraphics[width=\textwidth]{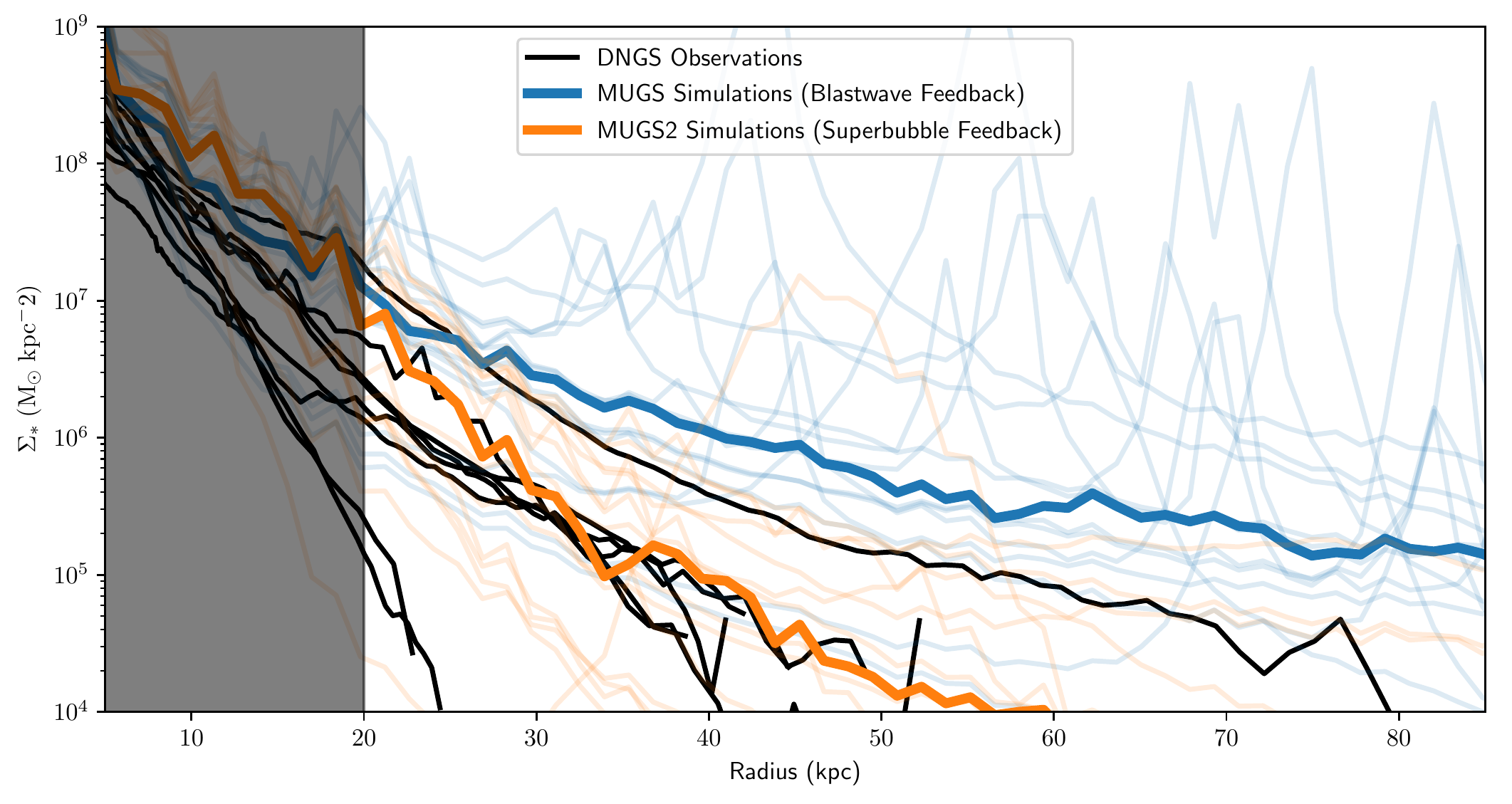}
    \caption{Stellar surface density profiles for the MUGS, MUGS2, and DNGS
    galaxies (blue, orange, and black curves).  The individual MUGS and MUGS2
    galaxies are shown as the thin, semi-transparent lines, while the median
    trend is shown by the thick solid line.  Despite having
    comparable stellar surface densities near the disk and in the inner halo
    ($<20\kpc$), the MUGS2 galaxies have stellar halos roughly 2 dex less dense
    beyond $\sim 50\kpc$ compared to the MUGS sample.  The shaded region
    indicates the inner parts of the stellar halo, while the unshaded region
    shows the outer stellar halo.}
    \label{sigma_star_all}
\end{figure*}
\begin{figure*}
    \includegraphics[width=\textwidth]{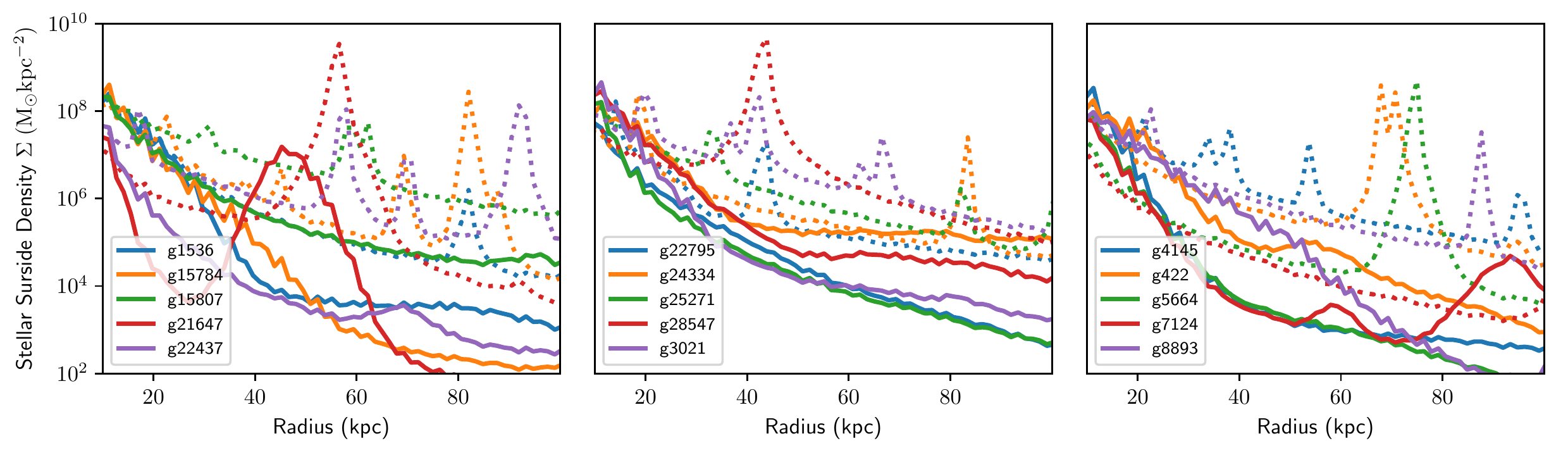}
    \caption{Comparison of stellar surface density in stellar halos of each pair
    of MUGS/MUGS2 galaxies.  The dotted curves show galaxies in MUGS, while the
    solid curves show galaxies in MUGS2.  We see significant diversity in the
    stellar halo surface densities in the MUGS/MUGS2 population, with >1 dex
    galaxy-to-galaxy variation in the outer stellar halo.  However, for all
    galaxy pairs, the MUGS2 outer stellar halos have much lower stellar surface
    density compared to their MUGS twin.}
    \label{sigma_star_pairs}
\end{figure*}

\begin{figure}
    \includegraphics[width=0.5\textwidth]{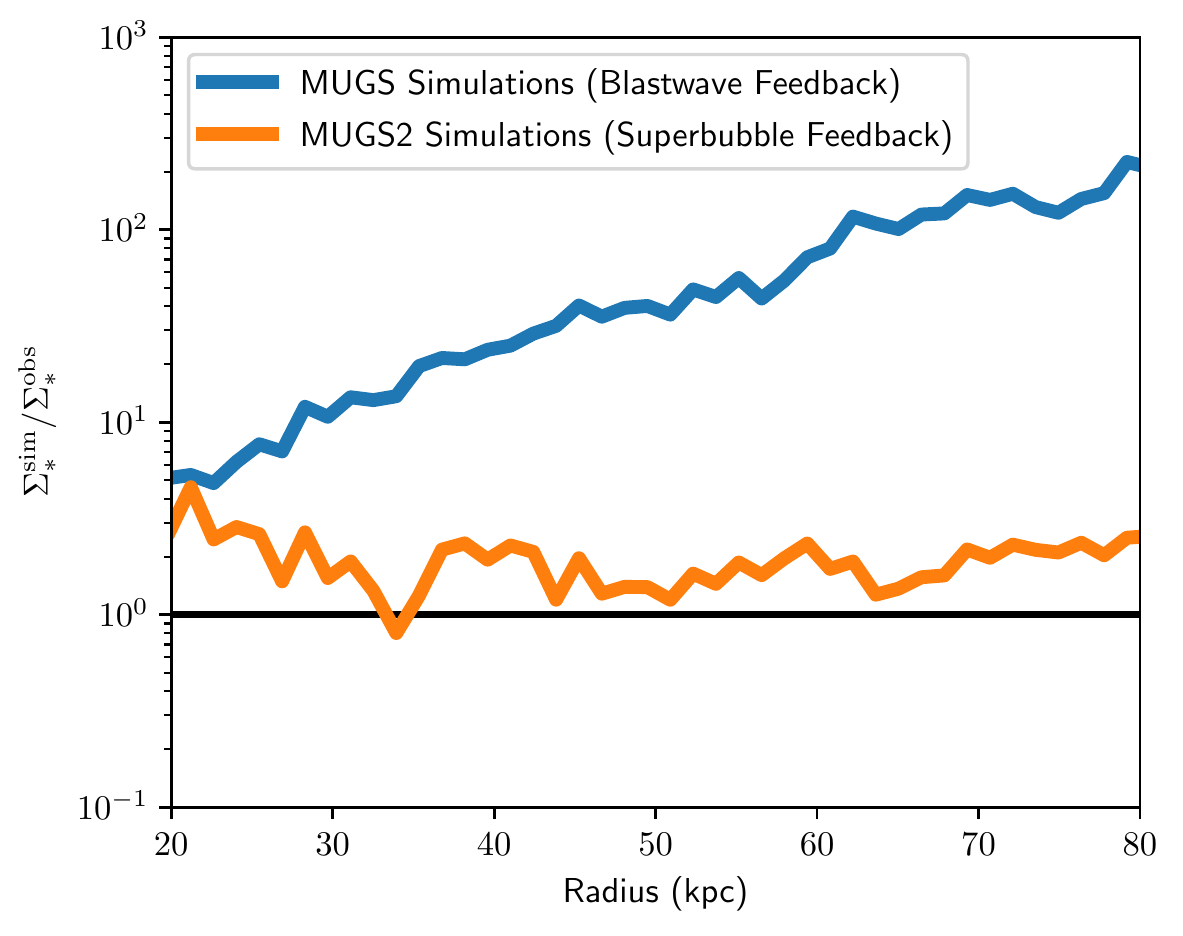}
    \caption{Ratio of median stellar surface density in the 15 MUGS/MUGS2
    simulations $(\Sigma_{\rm sim})$ to median stellar surface density observed
    from 8 DNGS galaxies $(\Sigma_{\rm obs})$ for the outer stellar halo from
    $20-80\kpc$.  The MUGS simulations have surface densities 1-2 dex higher
    than observed stellar halo outskirts (comparable to the outskirts of the
    Illustris-TNG100 galaxies in \citealt{Merritt2020}), despite having slightly
    lower overall stellar masses.  The MUGS2 stellar outskirts match the typical
    of surface densities in the DNGS outskirts.}
    \label{sigma_ratio}
\end{figure}

\begin{figure}
    \includegraphics[width=0.5\textwidth]{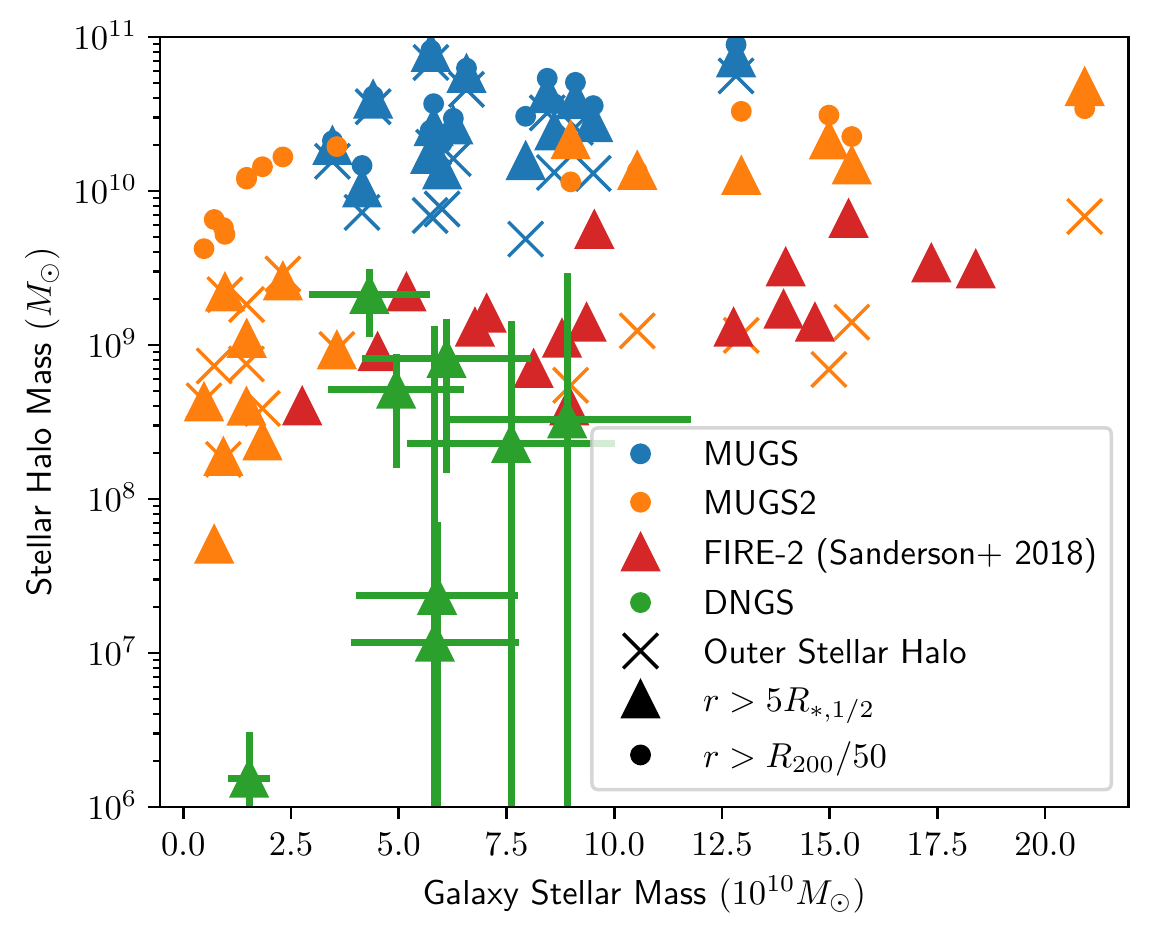}
    \caption{Stellar halo mass versus total stellar mass in the inner $30\kpc$
    for three different definitions of the stellar halo.  In addition to the
    stellar halo masses from MUGS and MUGS2, we also show the results from DNGS
    and the FIRE-2 simulations. We show a simple
    $r>20\kpc$ cut, measuring the outer stellar halo mass marked with a cross, a
    $r>5R_{*,1/2}$ cut marked with triangles (this is the definition of the
    stellar halo mass used in the \citealt{Merritt2016} observations shown in
    green, and from the \citealt{Sanderson2018} analysis of the FIRE-2
    simulations in red), and a $r>R_{200}/50$ cut marked with circles.  With selection
    criteria that include stars within $20\kpc$, we see smaller differences
    between our two samples of simulated galaxies (both $5R_{*,1/2}<20\kpc$ and
    $R_{200}/50<20\kpc$ for the MUGS and MUGS2 sample of simulated galaxies).
    As the stellar halos of the MUGS2 galaxies have steeper inner density
    profiles, most of their mass is contained within the inner regions
    $r<20\kpc$ of the halo.  The difference in the outer stellar halo matches
    what we would expect from the ratio in surface densities shown in
    Figure~\ref{sigma_ratio}.}
    \label{stellar_vs_halo}
\end{figure}

The galaxies in the MUGS/MUGS2 sample roughly span the range of stellar masses
probed by DNGS, as is shown in Figure~\ref{SMHMR}. We also show the simulated
galaxies studied by \citet{Sanderson2018} from the FIRE-2 simulations
\citep{Hopkins2018b}, which span a similar range of stellar and halo masses, and
the results from Illustris-TNG100 in the 2D histogram.  MUGS, MUGS2, and FIRE-2
all overproduce stars relative to the \citet{Behroozi2013} abundance-matched
stellar mass to halo mass relation above halo masses of $\sim10^{12}\Msun$,
likely owing to the absence of active galactic nucleus (AGN) feedback in all
three simulation suites (see \citealt{Keller2016} for a discussion of this in
the context of the MUGS2 simulations).  In the MUGS2 sample, we see that the
transition from ``regulated'' to ``runaway'' at $\sim10^{12}\Msun$ occurs rather
sharply (and somewhat stochastically), with some galaxies in this mass range
(g3021, g28547, and g24334) falling within the abundance-matched relation, while
others at slightly lower mass (g22437 and g22795) overproduce stars.  This is
due to a combination of the different assembly histories of these galaxies, halo
properties (g22437 and g22795 have the lowest spin parameters of the entire MUGS
sample, as seen in Table 1 of \citet{Keller2016}), and the stochastic effects
shown in \citet{Keller2019} and \citet{Genel2019}.  The SMHMR for TNG100, on the
other hand, has been specifically tuned to match the abundance-matched relation
of \citet{Behroozi2013} (see Figure 4 of \citealt{Pillepich2018}), and thus
tracks this relation extremely well.  With these samples, we can examine the
distinct evolution of the stellar halo in galaxies which overproduce stars in
the central disk and bulge (MUGS, MUGS2 above $M_{200}\sim10^{12}\Msun$, and
FIRE-2), and galaxies which match the abundance matched SMHMR (MUGS2 below
$M_{200}\sim10^{12}\Msun$ and TNG100).  As this shows, the significantly more
massive stellar halos in TNG100 are not a result of an overall overproduction of
stars: TNG100 by construction matches the abundance-matched SMHMR extremely well
(as was discussed in \citealt{Pillepich2018} and \citealt{Engler2021}.  In the
following sections we will see why this is not the case for the MUGS2 sample:
galaxies above and below $M_{200}=10^{12}\Msun$ both have significantly less
massive stellar halos than their MUGS counterparts.

A similar discrepancy between the simulated galaxies and observations can be
seen in the size-mass relation shown in Figure~\ref{size_mass}.  In the MUGS2
sample, galaxies with stellar mass below $\sim5\times10^{10}\Msun$ fall within the
range of observed stellar half-mass radii $R_{*,1/2}$ from \citet{Shen2003} and
\citet{Baldry2012}, while galaxies above this mass range all have values of
$R_{*,1/2}$ significantly lower than the mean observed values from \citet{Shen2003}.
The cause of this is the runaway growth of bulge mass that is discussed at
length in section 5.2 of \citet{Keller2016}, an effect that \citet{Dubois2015}
links directly to the onset of AGN feedback.  In the ``runaway'' MUGS2 galaxies,
the ultimate reason that both the stellar masses in Figure~\ref{SMHMR} and the
sizes in Figure~\ref{size_mass} begin to deviate from the observed relation is
star formation occurring primarily in the central bulge.

\subsection{The Impact of Feedback in Outer Stellar halos}
The stellar surface density as a function of radius for the DNGS sample, as well
as MUGS and MUGS2 is shown in Figure~\ref{sigma_star_all}.  As can be seen,
in all three samples of observed/simulated galaxies there is significant
galaxy-to-galaxy variance in the surface density of the outer stellar halo.  
Comparing these results to Figure 6 of \citet{Merritt2020}, the stellar halos of
the MUGS galaxies are qualitatively similar to those of the TNG100 sample.

Figure~\ref{sigma_star_pairs} shows the stellar halo in each of the 15 pairs of
galaxies from MUGS and MUGS2.  Peaks in these profiles correspond to satellites,
and as is clear, there are many more massive satellites in the MUGS simulations
compared to the MUGS2 simulations.  However, the diffuse halo between these
satellites also shows, in every galaxy, a markedly lower density in the MUGS2
simulations.  This is evidence that the satellites we see in MUGS have not
simply accreted with equal mass and then been stripped in the MUGS2 halos, but
have accreted with a much lower stellar in MUGS2.  The diversity of stellar
halos we see in each of these galaxies is a relic of the diverse assembly
histories of these simulated L* galaxies.  Notably, as we show in
\citet{Keller2015} and \citet{Keller2016}, superbubble feedback preferentially
suppresses star formation at high redshift and in low-mass galaxies, which we
examine later here in section~\ref{how_feedback_shapes}.  

The quantitative difference between the outer stellar halos in the MUGS and
MUGS2 simulations is shown clearly in Figure~\ref{sigma_ratio}.  Here, we
compare the ratio of the median outer stellar halo surface density from the two
simulation suites to the median outer stellar halo surface density from the DNGS
observations out to $80\kpc$ (the edge of the DNGS sample).  For the DNGS
observations, we fit each of the measured surface density profiles in the outer
halo (from $20-80\kpc$) with a power law, and take the median value of the 8
power-law fits to generate $\Sigma_*^{\rm obs}(r)$.  We apply this fitting
because the observed stellar surface density profile does not extend to $80\kpc$
for all 8 galaxies (only NGC4258 is measured out to $80\kpc$, owing to the
$31\rm{mag\; arcsec}^{-2}$ sensitivity limit of DNGS), and because the profiles have different
radii bins.  As Figure~\ref{sigma_ratio} shows, the surface density profile of
the median outer stellar halo in the MUGS2 galaxies matches the median observed
outer stellar halo surface density profile, to within a factor of $<2$ out to
$80\kpc$.  The MUGS galaxies, on the other hand, have stellar surface densities
$\sim5$ times higher than DNGS at $20\kpc$, increasing to $>100$ times higher
at $80\kpc$.  Even for the galaxies containing comparable total
stellar mass, the outer stellar halos have surface densities $1-2$
dex higher in MUGS compared to both MUGS2 and the DNGS observations, as is seen
in Figure~\ref{sigma_star_pairs}.

The total mass within the stellar halo is difficult to measure in a
self-consistent way for both observations and theory (see
\citealt{Sanderson2018} for a thorough discussion of this topic).  The critical
issue is that the strong negative slope of the stellar surface density means the
choice of inner radius for the stellar halo will have significant impact on the
stellar halo mass and the halo fraction.  It is especially important to ensure
that the inner radius of a region used to select halo stars is large enough that
it excludes the outer edge of the galaxy's stellar disk. As the gas disk can
extend much further than the stellar half-light radius $R_h$ (or even $5R_h$),
small amounts of star formation in the edge of the disk can contaminate the
population of halo stars with rotationally-supported disk stars.  In addition,
using the stellar half-mass radius will change the inner edge of the stellar
halo for galaxies of equal stellar/halo mass depending on their bulge fraction.
\citet{Merritt2020} identifies a tension between simulations and observations in
the stellar halo outskirts, beyond $20\kpc$, where these issues are largely
avoided.  With a fixed physical radius (roughly $0.1R_{200}$ for the halo masses
we examine here), we avoid both contamination from disk and bulge stars as well
as the bias introduced by measuring the halos from different inner radii.  Using
a fixed radius is particularly appropriate for this comparison, as each pair of
galaxies from the MUGS and MUGS2 simulations share ICs, and therefore should
have nearly identical DM accretion histories.  In Figure~\ref{stellar_vs_halo},
we reproduce the finding of \citet{Sanderson2018} and \citet{Merritt2020}
showing the sensitivity of the measured stellar halo mass to the choice of inner
radius.  The outer stellar halo (shown as crosses) reflects the large difference
in stellar surface densities we show in Figure~\ref{sigma_ratio}, while cuts
based on the stellar half-mass radius ($5R_{*,1/2}$, as is used in
\citealt{Merritt2016}) or a small fraction of the virial radius ($R_{200}/50$,
as is used in \citealt{Pillepich2014}) show a much more modest reduction in
stellar halo mass for the simulated galaxies with stellar masses above
$\sim5\times10^{10}\Msun$.  A major part of this arises not from a change in the
stellar halo, but from the reduced stellar half-mass radius that occurs in these
larger galaxies: as they become bulge-dominated, the half-mass radius is
significantly reduced.  As can be seen, measuring the outer halo mass from a
constant radius of $20\kpc$ gives reasonable agreement between the MUGS2
simulated stellar halos and the \citet{Merritt2016} Dragonfly observations. The
smallest value of $5R_{*,1/2}$ used in \citet{Merritt2016} is $10.4\kpc$,
compared to $3.4\kpc$ for \textsc{g22437}, the most bulge-dominated galaxy in
MUGS2.
These results are in good agreement with Figure 8 of
\citet{Merritt2020}, which also shows that the difference between the observed
and simulated stellar halo outskirts are much larger than the difference between
the stellar halo mass measured from a smaller radius ($2R_{*,1/2}$ in
\citealt{Merritt2020}).

\subsection{How Feedback Shapes Outer Stellar halos}
\label{how_feedback_shapes}
\begin{figure}
    \includegraphics[width=0.5\textwidth]{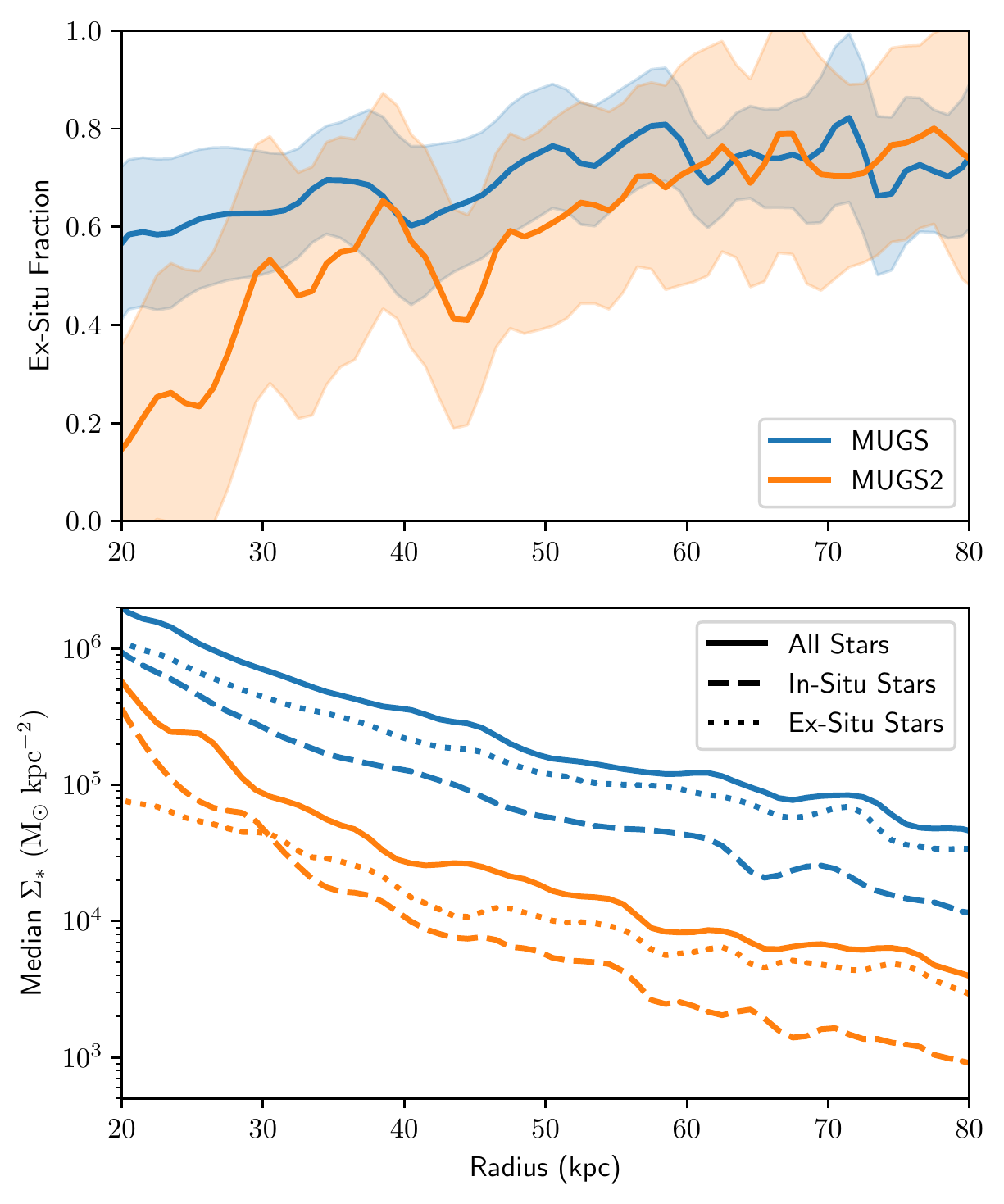}
    \caption{The origins of the outer stellar halos in MUGS and MUGS2.  In the
    top panel, we show the median ex-situ fraction in the MUGS and
    MUGS2 galaxies (blue and orange curves respectively). The standard deviation
    of each quantity is shown by the transparent areas. In the bottom panel, we
    show the median stellar surface density for all stars, ex-situ stars, and in-situ
    stars (solid, dashed, and dotted curves respectively).  Both MUGS and MUGS2
    show  ex-situ fractions above 0.5 for $r>40\kpc$.  In the MUGS sample, this
    fraction is roughly constant for the entire outer stellar halo, while the
    in-situ component of the MUGS2 outer stellar halo dominates for $r<40\kpc$.
    In the bottom panel we see that the ex-situ surface density exhibits a roughly constant
    reduction from MUGS to MUGS2, while the in-situ surface density is closer
    towards the inner fraction of the stellar halo outskirts.}
    \label{exsitu_fraction}
\end{figure}

\begin{figure}
    \includegraphics[width=0.5\textwidth]{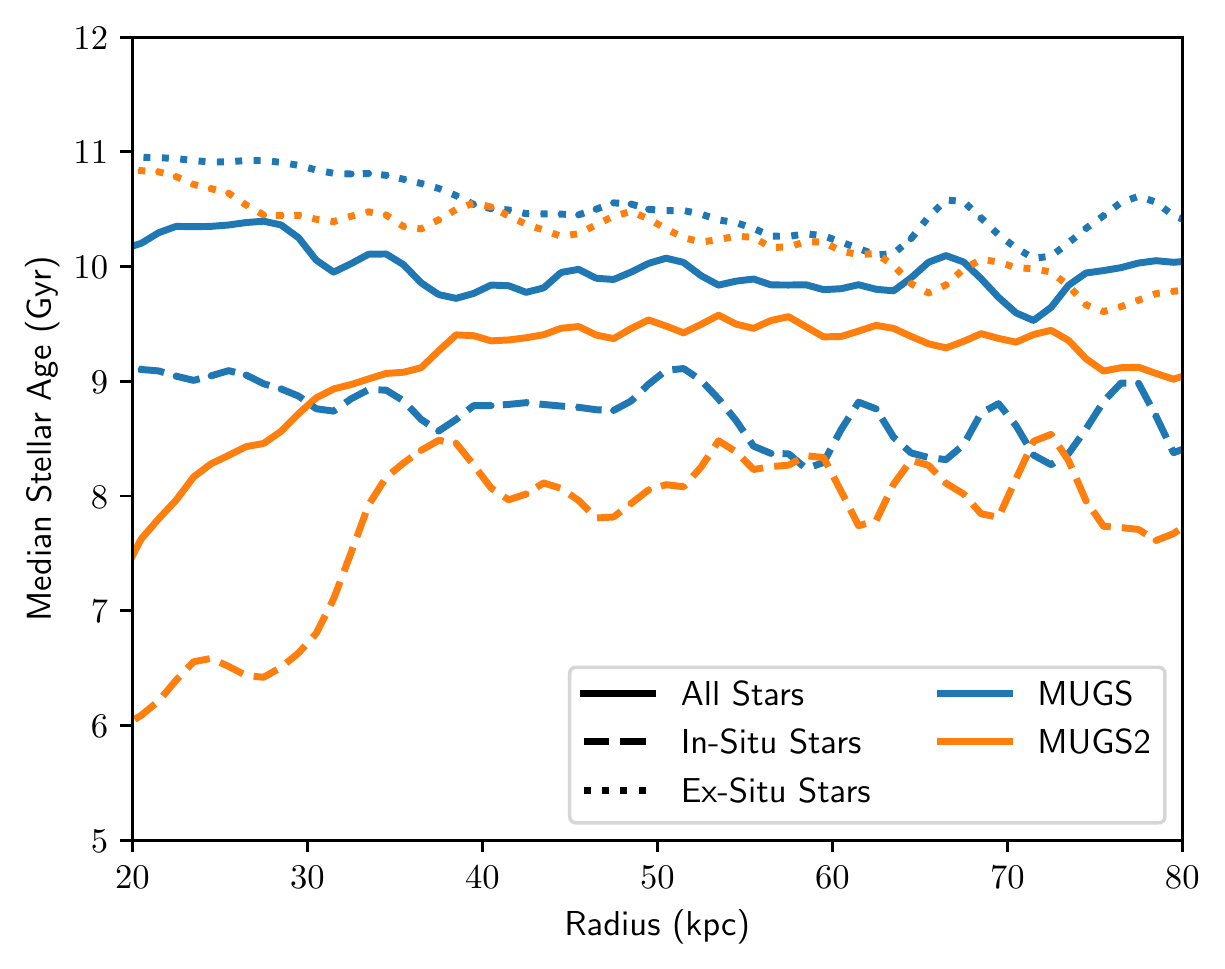}
    \caption{Median stellar age profiles for stars in the MUGS/MUGS2 simulations
    (blue/orange curves).  The median ages of all stars are shown by the solid
    curve, while in-situ stars are shown in the dashed line and ex-situ stars
    are shown in the dotted line.  The median age of ex-situ
    stars are higher than in-situ stars, and are relatively constant as a
    function of galactocentric radius.  The in-situ stars (which dominate the
    total stellar mass) are older in MUGS compared to MUGS2, and are relatively
    constant as a function of radius in MUGS.}
    \label{exsitu_age}
\end{figure}

\begin{figure}
    \includegraphics[width=0.5\textwidth]{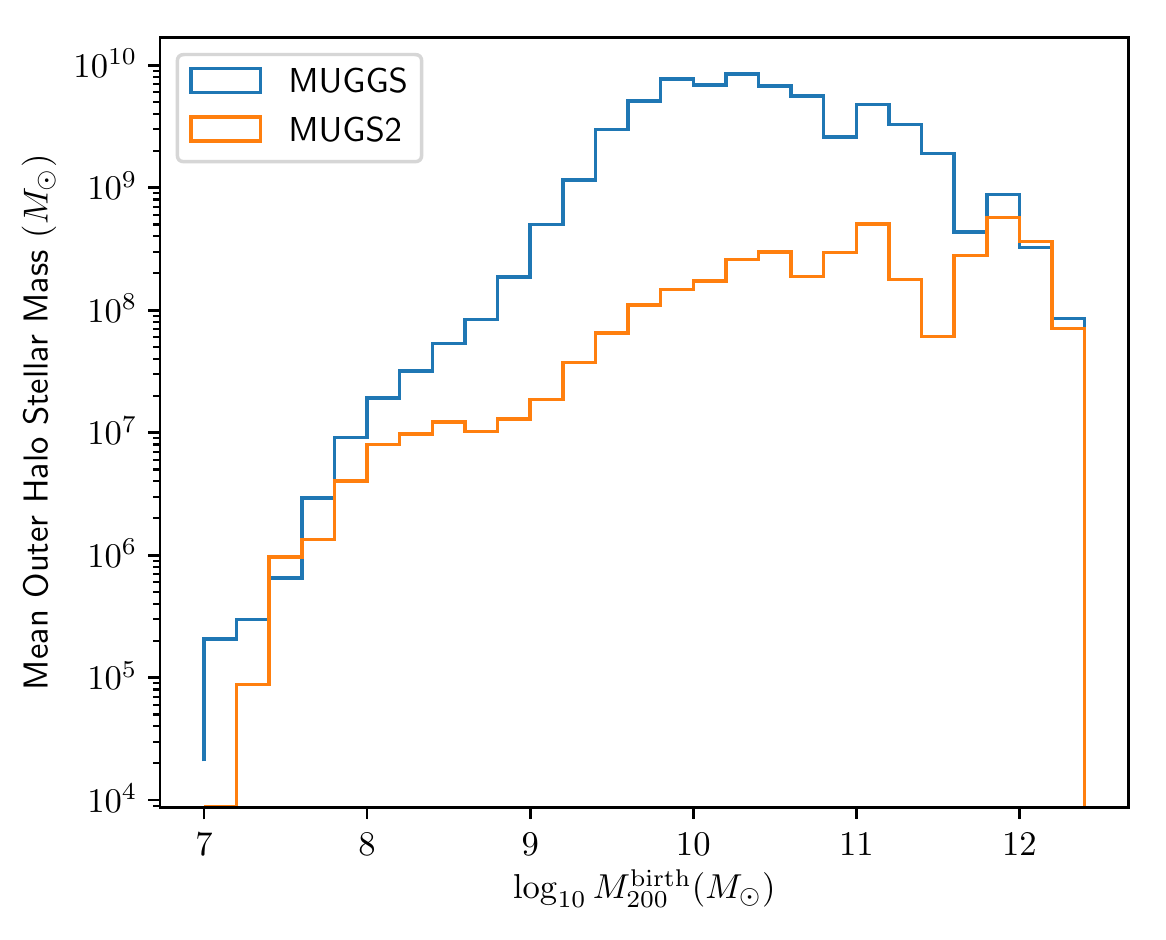}
    \caption{Mass as a function of birth halo mass $M_{200}$ for stars found in
    the outer stellar halo  $r>20\kpc$ for the 15 galaxies in the MUGS (blue) and
    MUGS2 (orange) samples. This figure shows the origin of the differences we
    have seen thus far in our simulated stellar halo outskirts.  Both simulation
    suites have roughly equivalent outer halo mass contributed from galaxies
    with $M_{200}^{\rm birth}>10^{11.5}\Msun$ and with $M_{200}^{\rm
    birth}<10^8\Msun$.  The ``superbubble'' feedback model used in the MUGS2
    simulations reduces the stellar mass entering the outer stellar halo that
    has formed in galaxies with a specific range of halo masses, from $M_{200}^{\rm
    birth}=10^8\Msun-10^{11.5}\Msun$, compared to the ``blastwave'' feedback
    model used in MUGS.}
    \label{halomass_birth}
\end{figure}

The results we have shown thus far demonstrate that the ``superbubble'' feedback
in MUGS2 produces significantly less massive outer stellar halos compared to the
``blastwave'' feedback in their equivalent MUGS simulations.  This effect
appears to be robust to the diverse assembly histories of the galaxies, and 
even whther or not the total stellar mass of the galaxy in MUGS and MUGS2
is similar.  This suggests that the outer stellar halo is a more sensitive
probe of stellar feedback than the stellar content of the disk (which dominates
the overall stellar mass).  In Figure~\ref{exsitu_fraction}, we show how the
different models for SN feedback impact the ex-situ and in-situ stellar
components of the two samples of simulated galaxies.  Previous studies (using
both semi-analytic models, e.g. \citealt{Bullock2005,Cooper2017} and
hydrodynamical simulations, e.g.  \citealt{Font2011,Pillepich2015}) have found
that anywhere from the majority to the totality of the stellar halo is composed
of tidal debris from accreted satellites (an ex-situ formation scenario).  

In the top panel of Figure~\ref{exsitu_fraction}, we show the radial profile and
$\pm1\sigma$ variance of the ex-situ fraction of stars in the MUGS and MUGS2
outer stellar halos as a function of galactocentric radius.  We define ex-situ
stars as those stars which are first identified in a halo other than the primary
progenitor of the galaxy, and are subsequently accreted by each galaxy.  The
ex-situ fraction is then the ratio of mass of ex-situ stars to the total stellar
mass (both measured within the outer stellar halo). As can
be seen in the top panel, the ex-situ fraction has a median of $\sim0.7$ in both
MUGS and MUGS2 galaxies beyond $\sim60\kpc$.   Inside $40\kpc$, the picture
becomes different for the two populations, with MUGS2 ex-situ fractions dropping
to $0.2$ at the beginning of the outer stellar halo.  This shows that much of
the difference we see is due to a reduction in overall contribution of ex-situ
stars, as even a complete elimination of in-situ stars could only reduce the
outer stellar halo surface density by at most a factor $25\%$ at $r>40\kpc$,
rather than the $>1$ dex reduction we see.  Of course, this also means that a
complete reduction of ex-situ stars cannot explain the differences we see.
Indeed, as we show in the bottom panel of Figure~\ref{exsitu_fraction}, the
surface density of both in-situ {\it and} ex-situ stars in the MUGS2 galaxies
are reduced by $>1$ dex beyond a galactocentric radius of $\sim40\kpc$.  As we
would expect from the top panel, the contribution of ex-situ stars are reduced
throughout the outer stellar halo in MUGS2.  This clearly shows that the lower
stellar halo surface densities in the MUGS2 galaxies arise from a reduction in
{\it both} in-situ and ex-situ halo stars.

Figure~\ref{exsitu_age} shows how the trends we see in the upper panels relate
to the median stellar age profiles of the MUGS and MUGS2 galaxies.  For both
in-situ and ex-situ stars, the outer stellar halo of both simulated galaxy
samples show only slight variation in in median stellar age as a function of
galactocentric radius.  As we would expect, the accreted ex-situ population is
older than the in-situ population for both the MUGS and MUGS2 outer stellar
halos.  Both the MUGS2 and MUGS ex-situ populations in the outer halo have
median ages of $\sim10\Gyr$, compared to a median age of $\sim6-8\Gyr$ for the
MUGS2 in-situ population and $\sim9\Gyr$ for the MUGS in-situ population.  The
population of in-situ stars most easily scattered into the outer halo by mergers
are those which formed at high redshift, when the merger rate was more frequent
and the disk potential was lower. 

The changes we see in the ex-situ fraction and stellar ages of the outer stellar
halos in Figure~\ref{exsitu_fraction} and Figure~\ref{exsitu_age} point towards
a difference in feedback efficiency in lower-mass galaxies.  Both ex-situ stars
and older in-situ stars are born in lower-mass galaxies than the central galaxy
at $z=0$.  In Figure~\ref{halomass_birth}, we show the birth galaxy halo mass
$M_{200}^{\rm birth}$ for all stars within the outer stellar halos of the MUGS
and MUGS2 galaxies.  We produce a single histogram for all 15 galaxies,
normalized by this amount to show the mean stellar mass within the outer stellar
halo as a function of birth galaxy mass.  Here we see clearly how the two
different SN feedback models in MUGS and MUGS2 produce outer stellar halos with
different masses, ex-situ fractions, and ages.  The ``superbubble'' feedback
model of MUGS2 reduces the contribution to the outer stellar halo from galaxies
with $M_{200}^{\rm birth}=10^8-10^{11.5}\Msun$.  Below this stellar mass range,
both feedback models are efficient, and a single $\sim10^5\Msun$ star particle
can halt the star formation of the progenitor (see \citealt{Mina2021} for a
recent comparison of these same feedback models in dwarf galaxies, albeit at a
somewhat higher galaxy mass range than $10^8\Msun$).  The outer stellar halo
contains very little mass from progenitors below $M_{200}^{birth}=10^8\Msun$.
Above $M_{200}^{\rm birth}= 10^{11.5}\Msun$, both feedback models become
inefficient, and SN will cease to regulate star formation.  This effect can be
seen in Figure~\ref{SMHMR}, and is discussed at length in \citep{Keller2016}.
The bulk of the mass in the MUGS outer stellar halos formed in galaxies with
$M_{200}^{\rm birth}=10^8-10^{11}\Msun$.  Superbubble feedback reduces the
contribution of from galaxies in this mass range to the outer stellar halo mass
by $1-2$ dex. For the galaxies with $M_{200}>10^{12}\Msun$, the total stellar
mass (and stellar mass in the inner $30\kpc$) is a weak discriminant of these
two feedback models.  By examining the outer stellar halo, we can probe the star
formation efficiency in lower mass galaxies, integrated over the assembly
history of the entire galaxy.

\section{Discussion \& Conclusion}
We have shown here how the modelling of SN feedback can have a dramatic effect
on the outer stellar halos of MW-mass galaxies.  We are only able to match the
observed stellar surface density profiles from the DNGS when SN feedback is
strong for galaxy halo masses $M_{200}=10^8-10^{11.5}\Msun$.  When SN feedback
is simulated with a physically-motivated superbubble model \citep{Keller2014},
which takes into account thermal conduction and evaporation and drives much more
efficient, buoyant galactic winds \citep{Keller2020a}, the outer stellar halo
surface densities are in agreement with the DNGS observations.  The MUGS2
galaxies, which include this new model, have outer stellar halos with
significantly reduced mass and surface density compared to the older MUGS
simulations, even when the galaxy mass becomes large enough that AGN feedback is
required to regulate the total stellar mass of the galaxy \citep{Keller2016}.
As the outer stellar halo is primarily composed of smaller accreted progenitors
and the tidal debris of interactions with the young primary progenitor galaxy,
the properties of stellar halo outskirts follow the early history of star
formation in the lower-mass progenitors of the final galaxy.

Despite having slightly higher total masses, the outer stellar halo of the MUGS2
galaxies show surface densities $1-2$ dex below those from the MUGS sample.  We
find that in both simulation samples, the most extended parts of the outer
stellar halos are dominated by old $(\sim10\Gyr)$ ex-situ stars (with ex-situ
fractions $\sim0.7$), in agreement with past simulation and semi-analytic studies
\citep[e.g.][]{Cooper2013,Tissera2012,Pillepich2015}.

\subsection{A Comparison of Feedback Models}
In this paper, we have primarily examined the impact on the stellar halos of
simulated, MW-mass galaxies from the MUGS and MUGS-2 simulations.  This work is
primarily motivated by the findings of \citet{Merritt2020} that simulated
stellar halos from Illustris TNG100 produce stellar halos with surface densities
10-100 times higher than observed stellar halos in DNGS.  A similar study
\citep{Sanderson2018} compared the total stellar halo mass from DNGS
observations to those from the FIRE-2 simulations.  Each of these simulation
projects relied on different feedback schemes to regulate the growth of stellar
and baryonic mass within their galaxies.  While TNG100 has been specifically
tuned to reproduce the \citet{Behroozi2013} SMHMR (see Figure~\ref{SMHMR}),
MUGS, MUGS2, and FIRE-2 instead have applied feedback models designed to
reproduce the small-scale behaviour of stellar feedback processes.  There are a
number of fundamental differences between these four models beyond this,
however.  

Of these four simulation projects (MUGS, MUGS2, FIRE-2, and TNG100), only TNG100
includes a model for AGN feedback, and only FIRE-2 includes a model for early
stellar feedback.  The model for AGN feedback in TNG100 seeds black holes in
halos that exceed a mass of $5\times10^{10}\Msun h^{-1}$.  This means AGN
feedback cannot occur in galaxies below a halo mass of $~7\times10^{10}\Msun$.
As we show in Figure~\ref{halomass_birth}, the majority of the additional halo
stars in MUGS are born in halos with masses near $10^{10}\Msun$, so a possible
explanation for the overly-massive stellar halos in TNG100 is that while AGN
feedback contributes significantly to regulating the stellar mass within the
inner $30\kpc$ of MW-mass galaxies, it is unable to regulate star formation in
the less massive halos where no black hole has been seeded.  The stellar feedback
model in TNG100 is an extension of the two-phase ISM model of
\citet{Springel2003}, along with an updated treatment for hydrodynamically
decoupled winds.  This model explicitly sets the velocity and direction of
winds, as well as their mass, energy, and metal loadings.  These quantities are
all determined by the local gas metallicity, DM velocity dispersion, and the
current Hubble constant $H(z)$.  The functional form setting these quantities,
as well as the dimensionless parameters may all contribute to the overproduction
of halo stars despite producing ``correct'' central galaxy stellar masses (for
example, the TNG wind velocity scales $v_w\propto H(z)^{-1/3}$, giving lower
outflow velocities at high redshift, when most halo stars are formed).

The stellar feedback models in MUGS, MUGS2, and FIRE-2 all produce outflows
emergently, with energy being injected on small scales and driving winds (or
not) as a function of the local ISM/CGM environment.  All three simulations
inject energy from core-collapse supernovae, with similar SN energies, and on
similar timescales \citep{Keller2022a}.  However, each relies on a different
method to prevent the overcooling of SN-heated gas.  MUGS simply delays
radiative cooling, using a method described in \citet{Stinson2006}.  MUGS2 uses
the superbubble method described in \citet{Keller2014} to put SN-heated gas into
a brief two-phase state, returning it to single-phase through thermal
evaporation of the cold phase into the hot.  FIRE-2 uses an updated mechanical
feedback \citep{Kimm2014,Hopkins2014} algorithm described in
\citet{Hopkins2018a} to partition SN energy into kinetic and thermal fractions
determined by the baryonic resolution, local gas density, and metallicity. This
algorithm injects the terminal momentum from the end of the pressure-driven
snowplow phase in gas with density and metallicity large enough that overcooling
will be significant for the resolution of the simulation.  In addition to the
different treatments for SN feedback, FIRE-2 also includes thermal feedback from
stellar winds, photoionization, and radiation pressure.  Each of these
components may contribute to regulating star formation in low-mass galaxies
differently than in high-mass galaxies.  A detailed comparison of the resolved
stellar halo outskirts in FIRE-2, building off of the work of \citet{Sanderson2018},
may be able to see if the same effects we observe in MUGS/MUGS2 occur with the
more comprehensive feedback models used in FIRE-2.  As we show in
Figure~\ref{stellar_vs_halo}, the FIRE-2 simulations also appear to have less
mass within their stellar halo than MUGS and TNG100, which suggests that the
feedback models in FIRE-2 and MUGS2 are both efficient at regulating star
formation in the low-mass progenitors of the outer stellar halo.

\subsection{Comparison to Other Simulations}
The goal of this study is to examine the findings shown in \citet{Merritt2020},
which is the first study to comprehensively compare the (non-integrated) stellar
halo profiles observed in \citet{Merritt2016} with simulations.  The authors of
this study a large discrepancy between the outer stellar halos of MW-mass
galaxies from the Illustris-TNG100 simulation \citep{Pillepich2018} and the DNGS
observations \citep{Merritt2016} of outer stellar halos (the ``missing
outskirts'' problem). The major difference between the simulations we examine
here and those of TNG100 are the larger population in TNG100.
\citet{Merritt2020} samples from a total of 1656 central galaxies with 
stellar mass $10^{10}-10^{11}\Msun$, selecting 50 galaxies chosen to match the
distribution of stellar mass in DNGS (compared to the 15 galaxy pairs we compare
here).  In addition to the larger sample size, TNG100 has somewhat lower
baryonic resolution than MUGS/MUGS2 ($2.2\times10^5\Msun$ in MUGS/MUGS2,
$1.4\times10^6\Msun$ in TNG100), as well as a significantly different model for
gas cooling, star formation, and feedback from both stars and AGN (see
\citealt{Pillepich2018} for details of the models in TNG100,
\citealt{Stinson2010} for the models in MUGS, and
\citealt{Keller2014,Keller2015} for the models used in MUGS2).  The results we
present here strongly suggest the differences between the outer stellar halos in
MUGS2, MUGS, and TNG100 are primarily a result of these different feedback
models, rather than the numerical resolution or sample size.

Of course, \citet{Merritt2020} is not the first study to examine stellar halos
through numerical simulation.  Both hydrodynamic and collisionless N-body
studies have attempted to trace the origin of the stellar halo in the MW and
other galaxies. The earliest attempts to study stellar halos through simulation
relied on collisionless N-body simulations
\citep[e.g.][]{Bullock2005,Abadi2006,Cooper2010,Cooper2013}.  While N-body
simulations are inexpensive compared to full hydrodynamic runs, they must use
particle tagging and semi-analytic models to translate the assembly history of
DM subhalos into a star formation history and stellar morphology.
\citet{Cooper2013} shows that the outer stellar halo is sensitive to the choice
of particle tagging parameters (in particular, the fraction of ``most bound'' DM
particles to tag as forming stars). The additional uncertainties introduced by
this process have been studied in \citet{Bailin2014}, which suggests they may
introduce significant systematics (though \citealt{Cooper2017} argue that this
effect can be mostly mitigated through more careful selection techniques for
particle tagging).

Hydrodynamic simulations of stellar halo formation have used both cosmological
zooms, as we have used here, as well as studies of large volumes (as is done in
\citealt{Merritt2020}).  Most of these focus on the entire stellar halo, with
either kinematic \citep[e.g.][]{Cooper2015}, spatial
\citep[e.g.][]{Pillepich2015}, or formation history-based
\citep[e.g.][]{Sanderson2018} cuts to divide the central galaxy from the stellar
halo.  As we have focused on the outer ($r>20\kpc$) stellar halo, and specifically its
surface density profile, these studies have delved much deeper into the
properties of stellar halos. One of the earliest studies, \citet{Zolotov2009},
used five SPH zoom-in simulations of MW-mass galaxies using the
same blastwave feedback model as MUGS.  Like in the MUGS/MUGS2 galaxies we have
studied, \citet{Zolotov2009} found that the in-situ component of the stellar
halo was concentrated towards the inner halo, and that the ex-situ fraction of
the total stellar halo was between $32-87\%$ (in contrast to the some of the
assumptions made in semi-analytic models such as \citealt{Bullock2005}).
\citet{Tissera2012} and \citet{Tissera2013} used six simulated MW zooms from the
Aquarius suite to study the chemical properties of halo stars based on their
binding energy, connecting this to their formation history.  More recently, the
latest generation of sub-grid modelling for star formation and feedback has
produced zoom-in simulations of MW galaxies with exceptional resolution and
agreement with observational scaling relations.  Simulations from FIRE
\citep{Hopkins2014,Sanderson2018}, the EAGLE-derived ARTEMIS \citep{Font2020}
and E-MOSAICS projects
\citep{Pfeffer2018,Kruijssen2019a,Hughes2019,Reina-Campos2020,Keller2020b}, and
the Illustris-derived AURIGA project \citep{Monachesi2019} have studied the
formation history, mass, and metallicity properties of stellar halos.  All of
these studies have found that accreted stars dominate the stellar halo,
especially at larger radii, and those which examine larger populations
\citep{Monachesi2019,Hughes2019,Sanderson2018} also find a large diversity in
the population of stellar halos of different MW-mass galaxies.  The E-MOSAICS
results \citep{Hughes2019,Reina-Campos2020,Keller2020b} also include a model for
globular cluster (GC) formation and evolution, and use these simulated GCs to
study the GC population in stellar halos, and compare them to the populations
of field stars in the galactic halo.  Future simulations that compare different
feedback models will allow us to study the co-evolution of stellar halos and
halo GCs in a similar fashion to the comparison we have presented here for the
overall stellar halos in MUGS and MUGS2. Large-volume simulations have also
been used to study the stellar halos of galaxies spanning a mass range much
wider than the narrow MW-mass region we have studied here
\citep{Font2011,Pillepich2014,Canas2020}.  While the sample size and mass ranges
probed by these larger-volume simulations exceed what we have shown here, they
all are limited by substantially lower resolution than what is capable in the
highest-resolution zoom-in simulations.  Studies adopting both approaches will
be essential to understand the full diversity of stellar halos and the impact
of feedback upon them.

\subsection{Limitations \& Future Work}
The primary limitations of this study are in numerical resolution and sample
size.  The MUGS and MUGS2 galaxies are cosmological zoom-in simulations with
baryonic mass resolution of $2.2\times10^5\Msun$.  While this is better than
what is achievable in the largest cosmological volumes, the highest-resolution
MW-mass zoom simulations are now achieving resolutions of $\sim10^4\Msun$
\citep[e.g.][]{Hopkins2018b}.  In the outer stellar halo, this resolution effect
can be significant.  At the lower surface density limits of surveys like DNGS,
$\sim10^4\Msun\kpc^{-2}$, our simulated stellar halos are relatively poorly
sampled, with a single star particle sampling a projected area of $10\kpc^2$.
Probing low-density areas will always be challenging for both observations and
simulations, but higher resolution will help to better sample the lowest-density
outskirts of the stellar halo and reduce the sampling noise that may occur in
lower-resolution studies.  The sample of galaxies we study here is also
relatively small, especially compared to the population from Illustris-TNG100
studied in \citet{Merritt2020}.  Given the diversity of outer stellar halo
profiles, this is a significant limitation that can only be overcome by
simulating larger volumes. Unfortunately, the observational samples are also
limited to small size, and so larger samples of simulated galaxies will not
allow us to make full statistical comparisons to observations in the near
future.

Our results have shown that, as has been seen in observations
\citep{Merritt2016,Harmsen2017} and simulations \citep{Merritt2020}, the outer
stellar halos of MW-mass galaxies exhibit tremendous diversity in their total
masses and surface density profiles.  We have limited our analysis to the impact
of SN feedback models on the stellar mass profiles of the outer stellar halo,
but there remains a great deal of information to be studied in the assembly
history of the individual MUGS/MUGS2 galaxies, as well as the metallicity and
kinematics of the stellar halo.  In a future paper (Khanom et al. in prep), we
will examine the differences in the satellite luminosity function of the MUGS
and MUGS2 galaxies, as well as the co-evolution of the stellar disk, bulge, and
halo in these matched sets of zoom-in simulations.  We leave as well study of
the metallicity and kinematics of their stellar halos to future work.

\subsection{Summary}
We have reproduced here the analysis of \citet{Merritt2020} which showed that
MW-like galaxies drawn from the Illustris-TNG100 simulations have significantly
more massive ``stellar halo outskirts'' (measured outwards from $20\kpc$)
relative to observed galaxies from the DNGS survey.  We use two sets (MUGS and
MUGS2) of 15 simulated galaxies with identical ICs and simulation code, but with
different models for stellar feedback by SNe to examine the sensitivity of
simulated stellar halo outskirts to the numerical implementation of stellar
feedback.  

We find that, even for galaxies with comparable total stellar mass,
simulated galaxies from MUGS2 produce outer stellar halos with surface densities
$1-2$ dex lower than simulated galaxies from the MUGS suite, in excellent
agreement with the DNGS observations.  We find that both the MUGS and MUGS2
simulations have stellar halo outskirts with large ex-situ fractions, in
agreement with past simulation studies.  The difference we find between the MUGS
and MUGS2 sample arises primarily from the more effective regulation of star
formation by stellar feedback in progenitor galaxies with
$M_{200}<10^{11.5}\Msun$.  These results show that the stellar halo outskirts of
massive galaxies are sensitive testbeds for stellar feedback, even when the
total stellar mass would be regulated by feedback from AGN.

This work shows that the ``missing outskirts'' problem identified by
\citet{Merritt2020} is sensitive to the details of how stellar feedback is
modelled in simulations of MW-like galaxies.  If a numerical model for SN
feedback is able to efficiently regulate star formation in the
$M_{200}<10^{11.5}\Msun$ progenitors of MW-like galaxies, the surface density
profiles of the outer stellar halos observed in DNGS can be reproduced. 

With growing observational capabilities and the increasing sophistication of
numerical simulations, stellar halo outskirts are a promising new frontier for
exploring the impact of galaxy assembly, star formation, and stellar feedback in
a cosmological environment.  

\section*{Acknowledgements}
BWK would like to thank Allison Merritt for useful discussions and for providing
data from DNGS. BWK would also like to thank James Wadsley, Sebastian
Trujillo-Gomez, Diederik Kruijssen, and Tessa Klettl for help editing and
proofreading this manuscript.   BWK gratefully acknowledges funding from the
European Research Council (ERC) under the European Union's Horizon 2020 research
and innovation programme via the ERC Starting Grant MUSTANG (grant agreement
number 714907).  

\appendix
\section{Robustness to Galaxy Orientation}
\label{orientation_robustness}
\begin{figure}
    \includegraphics[width=0.5\textwidth]{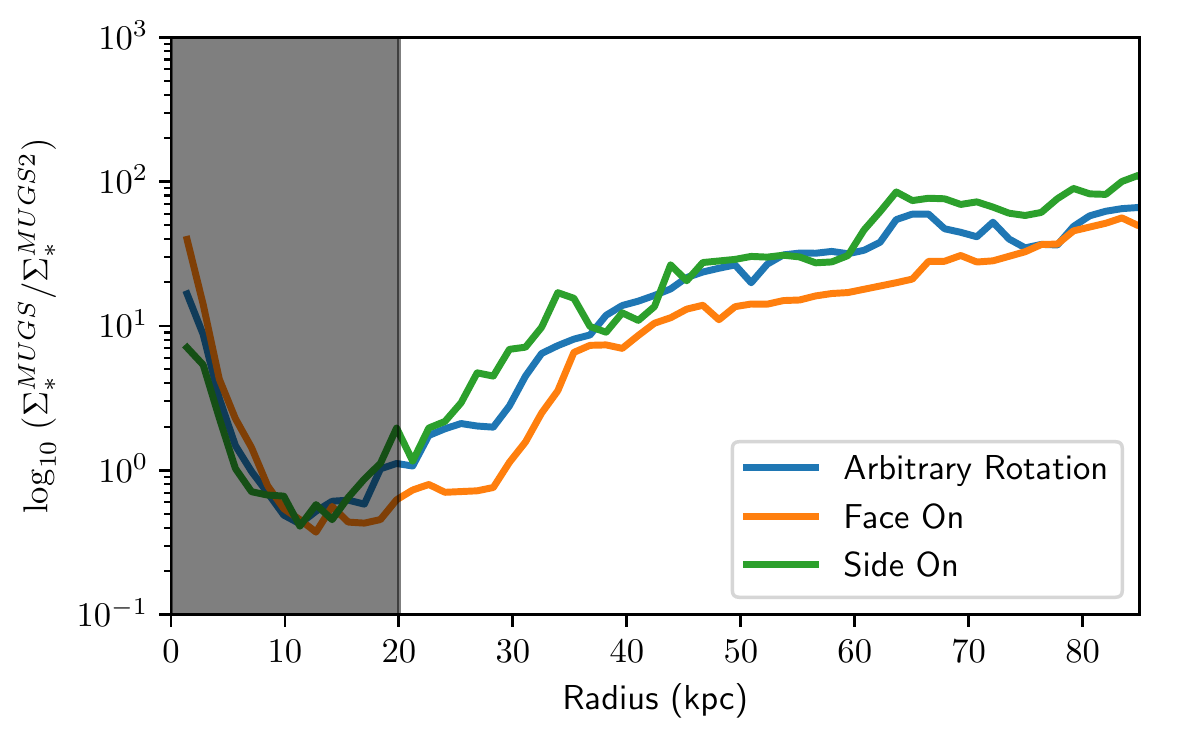}
    \caption{Relative surface densities of the MUGS and MUGS2 in random,
    face-on, and edge-on orientation.  Each curve is the ratio of the median
    surface density for each galaxy in MUGS and MUGS2.  As can clearly be seen,
    there is little impact in the relative surface density of the MUGS and MUGS2
    galaxies whether they are in a random, face-on, or edge-on orientation.}
    \label{orientation}
\end{figure}
Unfortunately for astrophysicists, galaxies are not spherically symmetrical.
The outskirts of a galaxy viewed edge-on will appear much brighter in surface
density than one viewed face-on, as we will be looking through a deeper column
of stars (of course, this also means that the solid angle covered by the
outskirts will be lower, as flux is still conserved).  As shown in
\citet{Elias2018}, this can increase the surface brightness in the outskirts by
up to 1.5 magnitudes (an issue also discussed in \citealt{Merritt2020}).  While
this will have the most impact in pencil-beam studies aligned along the major
and minor axes of the galaxies, it may be of some import here as well.  If, for
example, the MUGS2 galaxies have more oblate outer stellar halos, orientation
will change their observed surface brightness more significantly than for their
MUGS equivalent.  We have thus repeated the experiment shown by generating mock
observed surface density maps for our galaxies rotated to be face-on and
edge-on.  As we show in Figure~\ref{orientation}, the choice of orientation has
little impact on the relative effects we study here, altering the relative
surface densities in the outer stellar halo by only $\sim2$.  Regardless of galaxy
orientation, the MUGS galaxies always have $\sim10-100$ times larger surface
densities in the outer stellar halo.

\bibliographystyle{aasjournal}
\bibliography{references}

\end{document}